\documentclass[useAMS,usenatbib]{mn2e}

\usepackage{graphicx,amssymb,amsmath}

\title[A synthetic approach to the Oosterhoff dichotomy]{A stellar population
synthesis approach to the Oosterhoff dichotomy}
\author[Sollima et al.]{A. Sollima$^{1}$\thanks{E-mail:
antonio.sollima@oabo.inaf.it}, S. Cassisi$^{2}$, G. Fiorentino$^{1,3}$, R. G.
Gratton$^{4}$\\
$^{1}$ INAF Osservatorio Astronomico di Bologna, via Ranzani 1, Bologna, 40127,
Italy\\
$^{2}$ INAF Osservatorio Astronomico di Collurania, via Maggini s.n., Teramo,
64100, Italy\\
$^{3}$ Dipartimento di Fisica e Astronomia, Universit{\'a} di Bologna, via
Ranzani 1, Bologna, 40127, Italy\\
$^{4}$ INAF Osservatorio Astronomico di Padova, vicolo dell'Osservatorio 5,
Padova, 35122, Italy}
\begin{document}


\pagerange{\pageref{firstpage}--\pageref{lastpage}} \pubyear{2014}

\maketitle

\label{firstpage}

\begin{abstract}
We use color-magnitude diagram synthesis together with theoretical relations
from non-linear pulsation models to approach the long-standing problem of the
Oosterhoff dichotomy related to the distribution of the mean periods of
fundamental RR Lyrae variables in globular clusters. 
By adopting the chemical composition determined from
spectroscopic observations and a criterion to account for the hysteresis
mechanism, we tuned age and mass-loss to simultaneously reproduce the
morphology of both the turn-off and the Horizontal Branch of a sample of 17
globular clusters of the Milky Way and of nearby dwarf galaxies in the crucial
metallicity range ($-1.9<[Fe/H]<-1.4$) where the Oostheroff transition is apparent.
We find that the Oosterhoff dichotomy among Galactic globular clusters is
naturally reproduced by models. The analysis of the relative impact of
the various involved parameters indicates that the main responsibles of the 
dichotomy are the peculiar distribution of clusters in the age-metallicity plane 
and the hysteresis. In particular, there is a clear connection between the
two main branches of the age-metallicity relation for Galactic globular clusters
and the Oosterhoff groups. The properties of clusters' RR Lyrae belonging to other 
Oostheroff groups (OoInt and OoIII) are instead not well reproduced. While for 
OoIII clusters a larger helium abundance for a fraction of the cluster's stars 
can reconcile the model prediction 
with observations, some other parameter affecting both the Horizontal Branch
morphology and the RR Lyrae periods is required to reproduce the behavior of
OoInt clusters.  
\end{abstract}


\section{Introduction}
\label{intro_sec}

One of the most long-standing puzzles in stellar astrophysics is represented by 
the observed dichotomy in the mean period of globular cluster (GC) RR Lyrae 
variables pulsating in their fundamental mode (RR0). This evidence, first discovery 
by Grosse (1932) and Hachenberg (1939), was studied by Oosterhoff (1939, 
1944) who noted that GCs can be divided in two groups according to the mean
period of their RR0 variables: a first group with
$\langle P_{ab}\rangle\sim0.55~d$ (OoI) and a second group of clusters with significantly 
longer mean periods $\langle P_{ab}\rangle\sim0.65~d$ (OoII). Kinman (1959) noticed that the
subdivision of the two groups correlates with the clusters' metal content with 
the OoI group constituted by relatively
metal-rich ($-1.6<[Fe/H]<-1.0$) clusters and the OoII group comprising the most 
metal-poor ones ($[Fe/H]<-1.6$). Clusters of the two groups have also
differences in their relative fraction of first overtone (RR1) pulsator and
different horizontal branch (HB) morphologies (Stobie 1971).
A third group (OoIII) populated by two very metal-rich clusters ($[Fe/H]\sim-0.45$; 
NGC6388 and NGC6441)
which exhibit a long $\langle P_{ab}\rangle\sim0.75~d$ has been later defined (Pritzl et al.
2000). Only few clusters have a mean RR0 
period in the intermediate range $0.58<\langle P_{ab}\rangle<0.62$ (the
so-called "Oosterhoff gap"). Although at the epoch of the original
discovery of the Oosterhoff dichotomy, the surveys of variable stars were limited to only 5 
GCs, its significance has been confirmed
in the subsequent years when large samples of RR0 have been observed in many GCs. 
The same behavior has been observed among field RR0 populating the Galactic halo
(Sandage 1982a; Suntzeff, Kinman \& Kraft 1991; Miceli et al. 2008; Szczygie{\l}, Pojma{\'n}ski \& Pilecki 2009).
For decades the true nature of the Oosterhoff dichotomy has been debated
and many solutions have been proposed (see Catelan 2009 for a recent review). 
The first attempts to interpret this evidence focused their attention on the
temperature of the edges of the instability strip (Schwarzschild 1957) and on 
the mean luminosity of the HB (Sandage 1957). Both the above
interpretations have been subsequently discarded on observational grounds
(Sandage 1969; Stobie 1971). In a long series of papers A. Sandage and
collaborators investigated
the dependence of the periods of RR0 at a given temperature on metallicity
(the so-called "period-shift"; Sandage, Katem \& Sandage 1981, Sandage 1981,
1982a, 1982b, 1990). They concluded that while the progression of
the average RR0 period is continuous with metallicity, the gap is caused by the
non-monotonic behavior of the HB morphology with metallicity.
The subsequent studies indicated as possible responsibles the
hysteresis mechanism i.e. the metallicity dependence of the transition between the
fundamental and the first-overtone pulsation mode (van Albada \& Baker 1973; Stellingwerf 1975;
Caputo, Tornambe \& Castellani 1978; Bono et al. 1995; Bono, Caputo \& Marconi 
1995; Castellani,
Caputo \& Castellani 2003), the off-Zero Age HB (ZAHB) 
evolution occurring in OoII clusters RR0 (Lee, Demarque \& Zinn 1990; Lee 1991; 
Clement \& Shelton 1999) and the age differences among the two main Oosterhoff groups 
(Lee \& Carney 1999; Gratton et al. 2010).
Another puzzling evidence came to light 
when the observations of RR0 variables in extra-Galactic GCs and in
nearby dwarf galaxies were available: while Galactic GCs tend to avoid the
Oosterhoff gap, dwarf galaxies and their GCs populate preferentially
the gap (Walker 1992; Siegel \& Majewski 2000; Cseresnjes 2001; Dall'Ora et al.
2003; Soszynski et al. 2003; Greco et al. 2007, 2009; Kinemuchi et al. 2008;
Kuehn et al. 2008; Fiorentino et al. 2012; Coppola et al. 2013; Garofalo et al. 2013; Cusano et al. 2013). 
This finding has a deep relevance also in the context of the Galaxy formation
scenario since it represents a serious challenge to the identification of the 
dwarf galaxies as the building blocks of the Milky Way (Catelan 2009).
The clusters belonging to the OoIII group (NGC6388 and NGC6441) deserve a 
special discussion: these clusters host a large number of RR Lyrae at odds
with other GCs with similar metallicity. They also show an extended HB 
morphology with a tilted shape (Rich et al. 1997).
All these peculiarities can be explained by assuming an enhanced helium
abundance for the RR Lyrae in these clusters (Sweigart \& Catelan 1998; Moehler \& Sweigart
2006; Busso et al. 2007; Caloi \& D'Antona 2007;
Yoon et al. 2008; Moretti et al. 2009) possibly linked to a self-enrichment process occurred in these
GCs.

In this context, the overall picture has been furthermore complicated by
the discovery that GCs host multiple stellar populations with different
ages and p-capture elements abundances (Carretta et al. 2009b; Piotto 2009). 
The superposition of multiple populations with differences in their chemistry 
(in particular He and C+N+O abundances) can affect the morphology of the HB
(Carretta et al. 2009b; Gratton et al. 2013, 2014; Milone et al. 2014) as
well as the properties of their RR Lyrae variables (Jang et al. 2014).

While the main processes producing the Oosterhoff dichotomy have been recognized, their
relative efficiencies have still not been 
quantified in a large sample of GCs. 

In this paper we compare the observed color-magnitude diagrams 
(CMDs) of a sample of 13 GCs belonging to the Milky Way and to 4 GCs belonging to
two nearby satellite galaxies (1 in the Large Magellanic Cloud and 3 in the 
Fornax dwarf spheroidal galaxy) with synthetic CMDs with the aim of reproducing the
pulsational properties of their RR0 variables. Previous analyses of this kind have
been already performed in individual GCs by several groups (Marconi et al.
2003; Castellani,
Castellani \& Cassisi 2005) although none
of them involved the analysis of a large sample of GCs and was directly
focused on the Oosterhoff dichotomy.
In Section 1 we introduce the Oostherhoff dichotomy and discuss some of its 
statistical properties. Sect. 2 is devoted to the description of the method we
use to generate synthetic HB and RR Lyrae populations. We show the obtained
results and investigate the relative impact of the different physical
processes in Sect. 3. We summarize and discuss our results in Sect. 4.

\section{Statistical considerations} 
\label{stat_sec}

\begin{figure}
 \includegraphics[width=8.7cm]{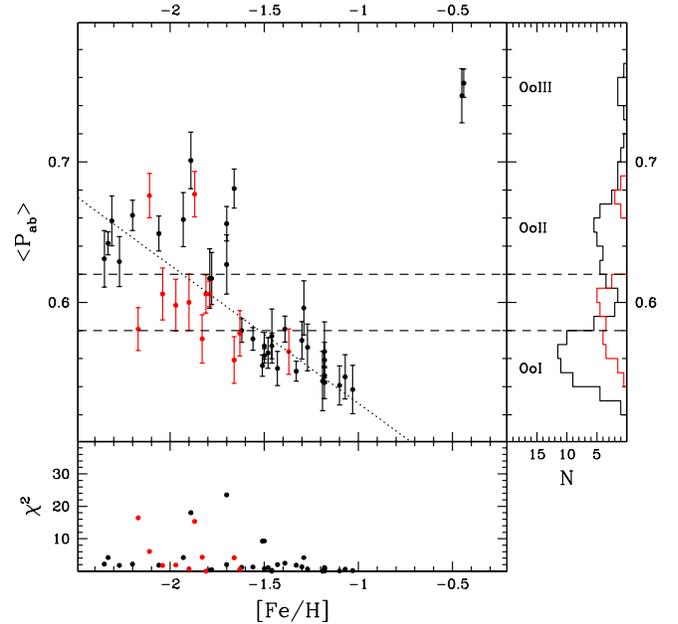}
 \caption{Mean RR0 period-metallicity diagram for Galactic GCs (black dots) and
 extra-Galactic GCs (red dots, grey in the printed version of the paper)
 containing at least 10 RR0. The boundaries of the Oosterhoff gap are marked by
 dashed lines. The linear fit to the data (excluding OoIII and extra-Galactic
 GCs) is indicated
 by the dotted line. The histogram of mean periods are shown in the
 right panel for both samples. In the bottom panel the individual $\chi^{2}$ of
 each GC with respect to the linear fit is shown.}
\label{oost}
\end{figure}

The pioneering work by Ooshterhoff (1939) was based on observations of just 5
GCs containing between 13 and 130 RR0 variables. In spite of the apparent
difference between the mean periods of the two identified groups, the sparse
number of clusters cast doubts on the significance of the dichotomy (see e.g.
{\'E}igenson \& Yatsyk 1985). During the following decades the census of RR Lyrae
variables in GCs has tremendously increased (Sawyer Hogg 1973; Clement et al.
2001) strengthening the evidence of the Oosterhoff dichotomy. In this section we
will analyse the statistical significance of the Oosterhoff dichotomy using the
last update of the Clement et al. (2001) catalog. In Fig. \ref{oost} the mean
periods of the RR0 variable are plotted as a function of the cluster
metallicity (from Carretta et al. 2009a). Only the 37 clusters with at least 10 RR0 have
been considered. The subdivision in the main Oosterhoff groups is apparent: a 
gap in the range $0.58<\langle P_{ab}\rangle <0.62$ almost devoid of clusters is
visible. 
For comparison, the GCs belonging to the Large Magellanic Clouds (LMC)
and to the Fornax dwarf spheroidal galaxy (from Mackey \& Gilmore 2003 and Greco
et al. 2007) are shown. As already noticed
by many authors (see Sect. \ref{intro_sec}) these GCs preferentially occupy the
Oosterhoff gap. The OoIII clusters NGC6388 and NGC6441 are also marked in this
diagram in a position which clearly stray from the mean RR0 period-metallicity
relation defined by the other clusters. 
It is interesting to note that {\it i)} there is no overlap in metallicity 
between the two groups i.e. all OoII GCs are more metal-poor than OoI ones,
although at $[Fe/H]\sim-1.65$ a number of GCs belonging to the two groups occupy
a small metallicity range, and {\it ii)} there is a well-defined linear trend of 
the mean RR0 period with metallicity: a Spearman rank test indicates a probability larger than 99.997\% that
the two variables are correlated. 
As a first step we estimate the probability that the distribution of mean
periods is bimodal. For this purpose we applied the KMM test for heteroscedastic
data (Ashman, Bird \& Zepf 1994) to the sample composed by the Galactic GCs excluding the OoIII
clusters. The test indicates a probability larger than 99.97\% that the periods
come from a bimodal distribution. 
As a second step we estimate the probability that the mean period of RR0
variables is a unique function of metallicity. For this purpose the $\chi^{2}$ test
has been applied to the same sample of clusters. The uncertainty of each measure has been estimated using the
error on the mean formula $\epsilon_{\langle P \rangle}=\sigma_{\langle P
\rangle}/\sqrt{N}$, where $\sigma_{\langle P\rangle}$ is the instrinsic spread
of the RR0 periods within each cluster and $N$ is the number of RR0.
The $\chi^{2}$ test gives a probability of less than $10^{-10}$ that the
$\langle P_{ab}\rangle$ distribution is drawn from the best-fit linear relation shown
in Fig. \ref{oost}. It is interesting to note that the largest contributors to
the $\chi^{2}$ are all GCs with a metallicity in the range
$-1.9<[Fe/H]<-1.5$ (see the bottom panel of Fig. \ref{oost}). By excluding these
clusters from the anlysis the $\chi^{2}$ probability increase to 13.8\%.
Summarizing, from the above test we state that the Oosterhoff dichotomy 
\begin{itemize}
\item{is statistically significant;}
\item{cannot be simply explained as a smooth trend of the mean RR0 period with
metallicity;} 
\item{it is
almost entirely determined by a small sample of GCs lying in a restricted range
in metallicity.}
\end{itemize}

\section{Method}
\label{met_sec}

\begin{table*}
 \centering
 \begin{minipage}{180mm}
  \caption{Summary of the performed simulations.}
  \begin{tabular}{@{}lcccccccccccc@{}}
  \hline
  NGC & Oosterhoff type & [Fe/H] & [$\alpha$/Fe] & HBR & $R_{GC}$ & Y & $t_{9}$ & $\Delta
  M$ & $\langle P_{ab}\rangle_{obs}$ & $\langle P_{ab}\rangle_{synth}$ &
  $F_{c,obs}$ & $F_{c,synth}$\\
      &                 &        &               &     &  kpc     &   & Gyr     & $M_{\odot}$ 
     & d                             & d & & \\
 \hline
 6229   & OoI   & -1.43 & +0.4 &  0.24 & 29.8 & 0.245 & 10.5 & 0.237 & 0.553$\pm0.012$ & 0.568 & 0.205$\pm0.080$ & 0.187\\	   
 6981   & OoI   & -1.48 & +0.4 &  0.14 & 12.9 & 0.245 & 11.0 & 0.216 & 0.564$\pm0.011$ & 0.558 & 0.159$\pm0.065$ & 0.215\\	    
 6584   & OoI   & -1.50 & +0.4 & -0.15 &  7.0 & 0.245 & 11.2 & 0.200 & 0.569$\pm0.010$ & 0.556 & 0.246$\pm0.069$ & 0.159\\
 5272   & OoI   & -1.50 & +0.4 &  0.08 & 12.0 & 0.245 & 11.5 & 0.202 & 0.562$\pm0.005$ & 0.556 & 0.220$\pm0.034$ & 0.225\\
 3201   & OoI   & -1.51 & +0.4 &  0.08 &  8.8 & 0.245 & 11.4 & 0.200 & 0.553$\pm0.008$ & 0.569 & 0.106$\pm0.037$ & 0.217\\
 6934   & OoI   & -1.56 & +0.4 &  0.25 & 12.8 & 0.245 & 10.7 & 0.215 & 0.574$\pm0.008$ & 0.574 & 0.127$\pm0.083$ & 0.289\\
 IC4499 & OoI   & -1.62 & +0.4 &  0.11 & 15.7 & 0.245 & 10.3 & 0.211 & 0.580$\pm0.008$ & 0.571 & 0.182$\pm0.053$ & 0.255\\
 7089   & OoII  & -1.66 & +0.4 &  0.96 & 10.4 & 0.245 & 12.3 & 0.222 & 0.681$\pm0.014$ & 0.698 & 0.395$\pm0.120$ & 0.371\\
 6656   & OoII  & -1.70 & +0.4 &  0.91 &  4.9 & 0.245 & 13.0 & 0.192 & 0.627$\pm0.021$ & 0.680 & 0.474$\pm0.192$ & 0.376\\
 5286   & OoII  & -1.70 & +0.4 &  0.74 &  8.9 & 0.245 & 12.0 & 0.190 & 0.656$\pm0.012$ & 0.621 & 0.434$\pm0.108$ & 0.340\\
 Rup106 & OoInt & -1.78 &  0.0 & -0.82 & 18.5 & 0.245 & 11.5 & 0.062 & 0.617$\pm0.018$ & 0.603 & 0.000$\pm0.000$ & 0.042\\
 Rup106 & OoInt & -1.78 & +0.4 & -0.82 & 18.5 & 0.245 & 10.2 & 0.140 & 0.617$\pm0.018$ & 0.590 & 0.000$\pm0.000$ & 0.062\\
 4833   & OoII  & -1.89 & +0.4 &  0.93 &  7.0 & 0.245 & 12.2 & 0.186 & 0.701$\pm0.020$ & 0.730 & 0.450$\pm0.181$ & 0.416\\
 6388   & OoIII & -0.45 &  0.0 & -0.69 &  3.1 & 0.260 & 12.5 & 0.305 & 0.676$\pm0.022$ & 0.468 & 0.550$\pm0.206$ & 0.140\\
 6388   & OoIII & -0.45 & +0.4 & -0.69 &  3.1 & 0.350 &  7.0 & 0.280 & 0.676$\pm0.022$ & 0.734 & 0.550$\pm0.206$ & 0.026\\
 1466   & OoInt & -2.17 &  0.0 &  0.42 &  --  & 0.245 & 13.2 & 0.035 & 0.591$\pm0.015$ & 0.640 & 0.405$\pm0.116$ & 0.687\\
 1466   & OoInt & -2.17 & +0.4 &  0.42 &  --  & 0.245 & 12.8 & 0.092 & 0.591$\pm0.015$ & 0.607 & 0.405$\pm0.116$ & 0.489\\
 F2     & OoI   & -1.83 &  0.0 &  0.50 &  --  & 0.245 & 12.1 & 0.115 & 0.574$\pm0.017$ & 0.618 & 0.302$\pm0.096$ & 0.513\\
 F2     & OoI   & -1.83 & +0.4 &  0.50 &  --  & 0.245 & 10.5 & 0.187 & 0.574$\pm0.017$ & 0.602 & 0.302$\pm0.096$ & 0.379\\
 F3     & OoInt & -2.04 &  0.0 &  0.44 &  --  & 0.245 & 11.5 & 0.093 & 0.606$\pm0.018$ & 0.642 & 0.384$\pm0.073$ & 0.643\\
 F3     & OoInt & -2.04 & +0.4 &  0.44 &  --  & 0.245 & 10.2 & 0.161 & 0.606$\pm0.018$ & 0.595 & 0.384$\pm0.073$ & 0.446\\
 F4     & OoInt & -1.90 &  0.0 & -0.42 &  --  & 0.245 & 12.0 & 0.045 & 0.600$\pm0.020$ & 0.607 & 0.120$\pm0.073$ & 0.081\\
 F4     & OoInt & -1.90 & +0.4 & -0.42 &  --  & 0.245 & 10.5 & 0.137 & 0.600$\pm0.020$ & 0.595 & 0.120$\pm0.073$ & 0.144\\
\hline
\end{tabular}
\end{minipage}
\end{table*}

\begin{figure}
 \includegraphics[width=8.7cm]{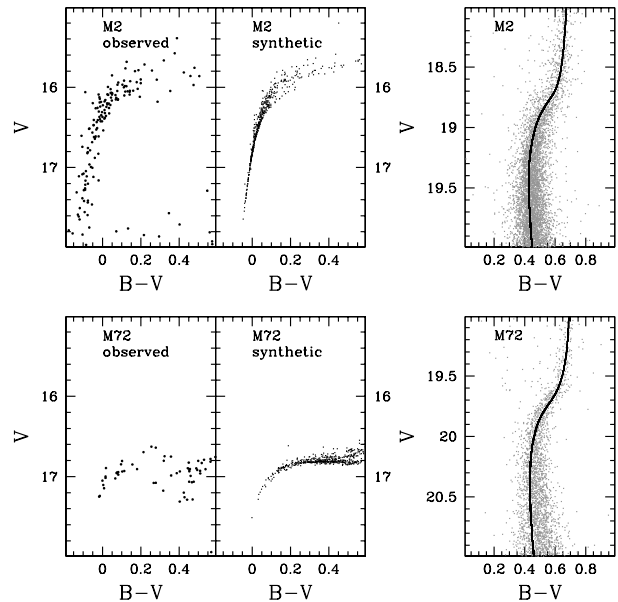}
 \caption{Best-fit of the SGB (right panels) and HB (left panels) morphology of
 the two sample clusters M72 (OoI; bottom panel) and M2 (OoII; top panel).}
\label{cmd}
\end{figure}

\begin{figure*}
 \includegraphics[width=13cm]{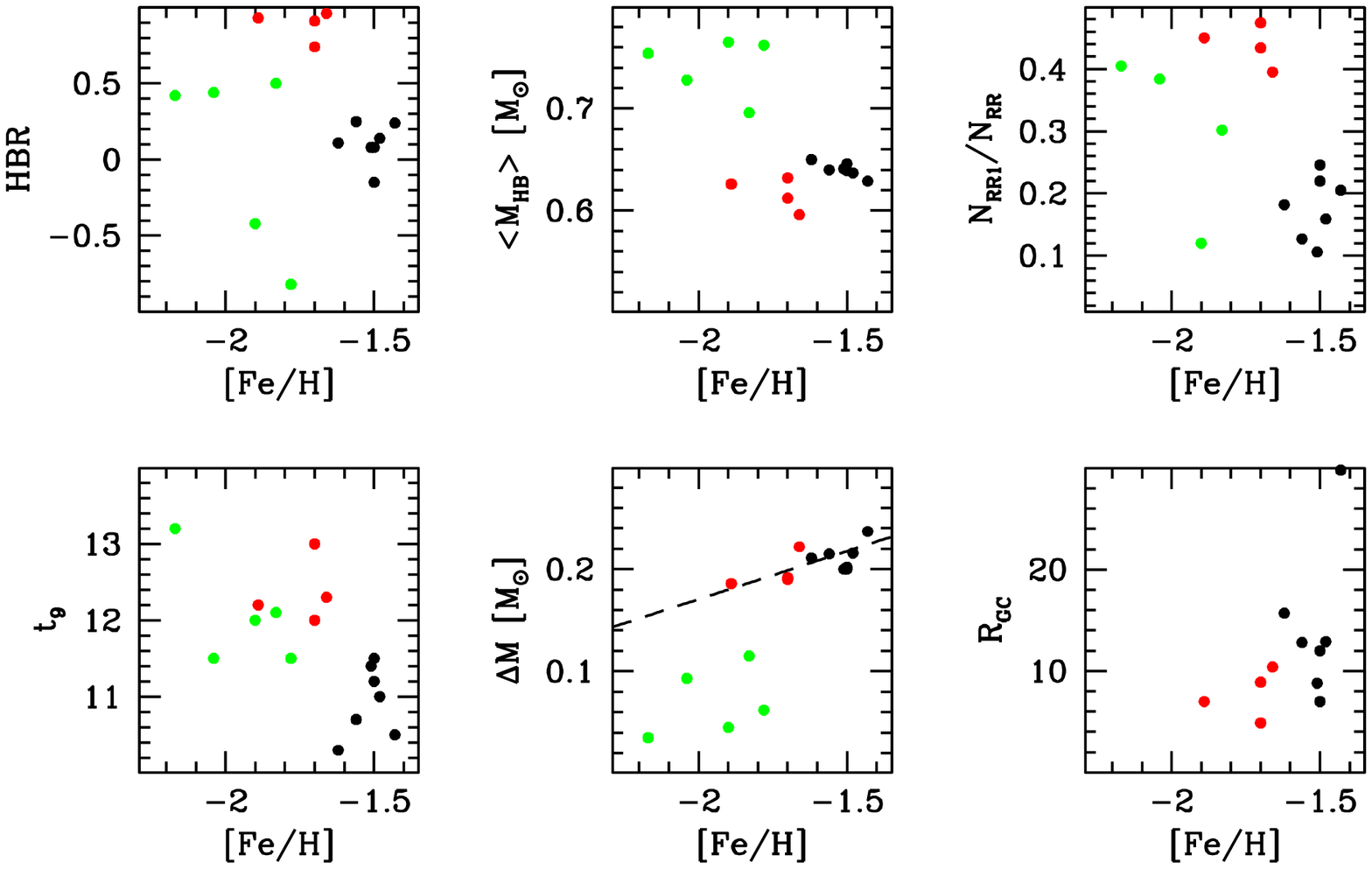}
 \caption{Best-fit ages (bottom-left panel), mass loss (bottom-central panel), 
 Galactocentric distance (bottom-right panel), HB
 type (upper-left panel), mean HB mass (upper-central panel) and fraction of RR1
 (upper-right panel) as a function of
 metallicity. Black, red and green dots (black filled, grey and black open in
 the printed version of the paper) represent Galactic OoI, OoII and
 extra-Galactic clusters, respectively. The mass loss-metallicity relation by 
 Gratton et al. (2010) is marked in the bottom-central panel with a dashed line.}
\label{corr}
\end{figure*}

We approached the problem by comparing the CMDs of a selected sample of GCs 
with a set of synthetic CMD.
On the basis of the conclusion reached in Sect. \ref{stat_sec}, we selected 
those clusters with at least 10 RR0 variables and a metallicity in
the range $-1.9<[Fe/H]<-1.4$ (in the Carretta et al. 2009a scale).
Twelve Galactic GCs are included in this sample, 7 belonging to the OoI group 
(namely NGC6229, M72, NGC6584, M3, NGC3201, NGC6934 and IC4499), 4 belonging to
the OoII group (M2, M22, NGC5286, NGC4833) and 1 to the OoInt group (Ruprecht
106). We also added to the sample one OoIII cluster (NGC6388) and the 4 
clusters belonging to nearby satellites of the Milky Way with more than 10 RR0
(NGC1466 in the LMC and clusters F2, F3 and F4 in the Fornax dSph).
We used the photometric catalogs provided by the snapshot survey (Piotto et al.
2002) which provided accurate BV photometry (converted from the F439W and F555W
HST magnitudes using the transformations procedure reported in Piotto et al.
2002) for most of the target GCs.
For those clusters not included in the above survey we used the
ground based photometry by Rosenberg et al. (2000a, 2000b; for M3 and M22),
Zorotovic et al. (2009; for NGC5286), Buonanno et al. (1990; for Rup 106),
Johnson et al. (1999; for NGC1466) and Buonanno et al. (1998, 1999; for the
Fornax clusters).   
For each cluster we computed a synthetic CMD by extracting a large number 
($N>10^{6}$) of stars from a Kroupa (2002) mass function and populating the
different regions of the CMD by interpolating within the set of
evolutionary tracks from the BASTI database (Pietrinferni et al.
2006)\footnote{http://www.oa-teramo.inaf.it/BASTI}. For Milky Way GCs we adopted 
the metallicity of Carretta et al. (2009a) and an $[\alpha/Fe]=+0.4$ which
appears to be a suitable choice for all GCs in the considered metallicity range
(Carretta et al. 2009b) but for the case of Rup106 for which a lower 
$\alpha$-enhancement $[\alpha/Fe]=0.0$ has been suggested by a recent
spectroscopic analysis (Villanova et al. 2013).
For the extra-Galactic GCs the metallicities reported by Catelan et al. (2009)
have been adopted. Although studies on field stars in nearby dSphs indicated a
$[\alpha/Fe]$ value smaller than that observed in the Milky Way (Venn et al.
2004; but see Lemasle et al. 2013), direct spectroscopic measures of these abundance patterns
are available for only two clusters of our sample (Letarte et al. 2006) and
suggest a $[\alpha/Fe]$ comparable with that of Galactic GCs. 
For this reason we performed simulations assuming for these GCs both a
solar-scaled and an $\alpha$-enhanced ($[\alpha/Fe]=+0.4$) mixture.
For a given choice of metallicity and $[\alpha/Fe]$
we distributed stars on the HB tracks by extracting their masses from a Gaussian
distribution with a given mean and spread (see below), and assuming that stars 
are being fed onto the HB at
a constant rate (Salaris, Cassisi \& Pietrinferni 2008). The bestfit has
been performed in an objective manner by comparing the same measurable
quantities in the observed and synthetic CMDs.
The age of each cluster has been tuned to
reproduce the distance between the mean locus of the HB and the level of the 
Sub Giant Branch (SGB) as measured in the observed CMDs. The derived ages turn
out to be in agreement with those found in previous studies: the mean age 
difference for the 8 GCs in common with the works by Mar{\'{\i}}n-Franch et al. (2009)
and VandenBerg et al. (2013) are $\Delta t_{9,MF}=0.05\pm0.58$ Gyr and $\Delta
t_{9,VdB}=0.24\pm0.54$, with
a r.m.s compatible with the combined uncertainties of the considered studies and
indicating no systematic offset.
Once the cluster age has been estimated the mean HB mass has been chosen to fit
the observed color distribution of HB stars and the HB ratio (HBR=(B-R)/(B+V+R)
where B, R are the number of stars bluer and redder than the instability 
strip and V is the number of RR Lyrae variables; Zinn 1986). To 
limit the number of free parameters, we adopted a fixed mass spread along the HB of
$\sigma_{M}=0.02~M_{\odot}$ (Lee, Demarque \& Zinn 1994). Fortunately, 
there is no covariance between the two parameters left free in the fitting
procedure ($t_{9}$ and 
$\langle M_{HB}\rangle$). This allows to perform the bestfit
of these parameters sequentially ensuring the absence of degeneration in the 
derived solution. The bestfit solutions range within $7.0<t_{9}/Gyr<13.2$ and
$0.58<\langle M_{HB}\rangle/M_{\odot}<0.78.$

We conducted our simulations assuming each GC as a single stellar
population. As already introduced in Sect. \ref{intro_sec}, it has been 
shown that GCs host multiple stellar populations showing different
light elements abundances (Carretta et al. 2009b; Piotto 2009). In principle,
the shape of the CMD as well as the properties of RR Lyrae variables could be
affected by a spread in the abundance of some of these elements (in particular
He and the total C+N+O abundance). In particular, Jang et al. (2014) pointed out
that the Oosterhoff dichotomy could be actually reproduced if the RR Lyrae
population of the two Oosterhoff groups belong to different stellar generations
characterized by a particular combination of ages, He and CNO abundances. 
However, large He differences 
($\Delta Y>0.05$) among the stellar populations of individual GCs have been suggested 
only for a few GCs: $\omega$ Cen, NGC2808 and NGC2419
(Norris 2004; Piotto et al. 2005; di Criscienzo et al. 2011; not included in this analysis) and the two OoIII GCs (see Sect. \ref{ooiii_sec}).
Moreover, because of their high effective temperature, He-rich stars are expected to 
populate the blue portion of the HB without contributing to the RR Lyrae 
population.
On the other hand, differences in the C+N+O abundance seem 
to be limited to a small sample
of GCs such as $\omega$ Cen, NGC1851 and M22 (Yong et al. 2009;
Villanova, Geisler \& Piotto 2010; Alves-Brito et al. 2012; Marino et al. 2012)
\footnote{Note that, since the effect of CNO abundance on the HB morphology is 
produced by the increase of the energy generation in the H-burning shell through
the CNO-cycle, the total C+N+O determines significant changes in the HB
evolution, while individual abundance variations of these elements at fixed 
C+N+O have a minor impact.}.This is consistent with the scenario where the
responsibles for the self-enrichment of GCs are massive stars not experiencing
the third dredge-up, thus preserving the total C+N+O abundance during the CNO 
cycle burning (D'Ercole et al. 2008).  
The fact that the selected GCs seem not to be affected by large He and CNO
enhancement between the various sub-populations should
guarantee that the average pulsational properties of the RR Lyrae populations 
are similar to those expected for a single stellar population with  
canonical abundances of these elements. Small (possible) differences in the 
He/CNO abundance among different stellar
populations are expected to produce a spread in the RR Lyrae properties 
mimicking a single population
cluster with a slightly different mean He and CNO abundance (see Gratton et al. 2011, 2012a,
2013, 2014).

In Fig. \ref{cmd} the best-fit solutions for two clusters (M2 and M72) are 
shown.

The pulsational properties of the RR Lyrae in each synthetic CMD have been
then calculated using the relations derived by the non-linear pulsation models of
Marconi et al. (2003). In particular we considered RR Lyrae variables all those
stars lying within the blue edge of the first-overtone pulsators (FOBE) and the red
edge of the fundamental pulsators (FRE) as defined in Marconi et
al. (2003; we adopted the relations provided for a mixing-length value of 
$l/H_{p}=2$). The other two borders have been defined by interpolating among the
loci provided by Bono, Caputo \& Marconi (1995). 
It is well known that there is a region of overlap between the
instability strips of fundamental and first-overtone pulsators (the so-called
"OR region") where both modes of pulsations can occur (van Albada \& Baker
1973). For the stars falling in this region we adopted the hysteresis criterion defined by van Albada \& Baker
(1973): a star falling in the OR is considered an RR0 if it
entered in the OR region from its red edge (FORE) while it is considered a
RR1 if it entered from the blue one (FBE). For the sample of
(few) stars which passed all their evolution in the OR region we considered
the cooler half of them RR0 and the hottest half RR1.
The period of the RR0 variables has been calculated then from the relation
of Marconi et al. (2003)
\begin{eqnarray}
\log{P/d}&=&11.066+0.832~\log{L/L_{\odot}}-0.650~\log{M/M_{\odot}}\nonumber\\
         & &-3.363~\log{T_{eff}/^{\circ}K}
\label{per_eq}
\end{eqnarray}
As already found in many previous studies (Cassisi et al. 2004; Castellani et
al. 2005; Salaris et al. 2013) the above approach gives mean periods which
are significantly longer than those observed in real GCs. This is probably due
to uncertainties in the complex mechanisms governing the mass size of 
the He core in the Red Giant progenitors (e.g. cooling by neutrinos, electron
conduction, etc.) and in the mixing-length determining the exact temperature of
the FRE. So, we applied a decrease of the HB luminosity by $\Delta
\log{L/L_{\odot}}=-0.03$. It is interesting to note that updated stellar
model computations based on more recent evaluations for the conductive opacity 
(Cassisi et al. 2007) predict a reduction of the 
HB luminosity  of $\Delta{\log(L/L_\odot)}\approx -0.02$ with respect the standard
BaSTI models. This correction is smaller than the typical
uncertainties due the above processes (Cassisi et al. 1998; Castellani \&
Degl'Innocenti 1999).
The whole set of simulations is summarized in Table 1.

\section{Results}
\label{res_sec}

\subsection{Milky Way GCs: Oosterhoff I and II}

\begin{figure}
 \includegraphics[width=8.7cm]{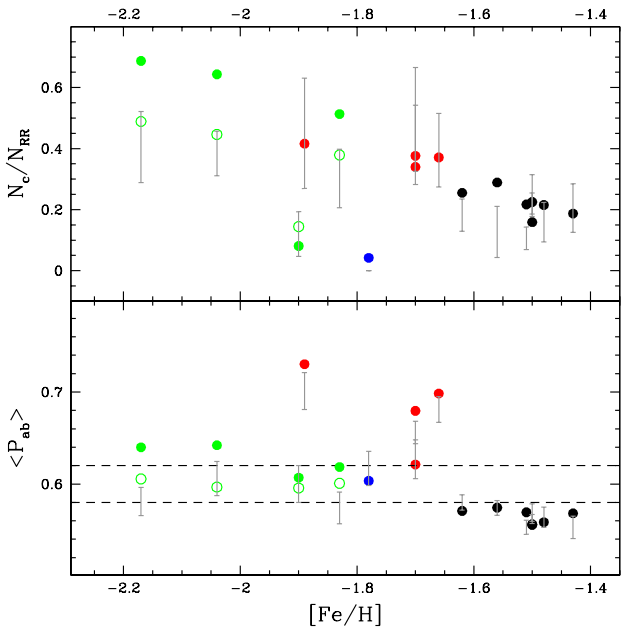}
 \caption{In the bottom panel the synthetic RR0 mean periods as a function of 
 metallicity are shown. In the top panel the relative fraction of RR1 as a
 function of metallicity are shown. Black filled points
 represent Galactic OoI GCs, red points (open dots in the printed version of the
 paper) represent Galactic OoII, blue points (open star in the printed version)
 represent Galactic OoInt GCs, 
 filled and open green points (grey in the printed version) represent extra-Galactic GCs with 
 [$\alpha$/Fe]=0.0 and [$\alpha$/Fe]=+0.4, respectively. The observed
 $\pm1~\sigma$ ranges in $\langle P_{ab}\rangle$ and $N_{c}/N_{RR}$
 are also indicated for comparison with grey errorbars. The dashed lines in the
 bottom panel mark the borders of the Oosterhoff gap.}
\label{res}
\end{figure}

\begin{figure*}
 \includegraphics[width=13cm]{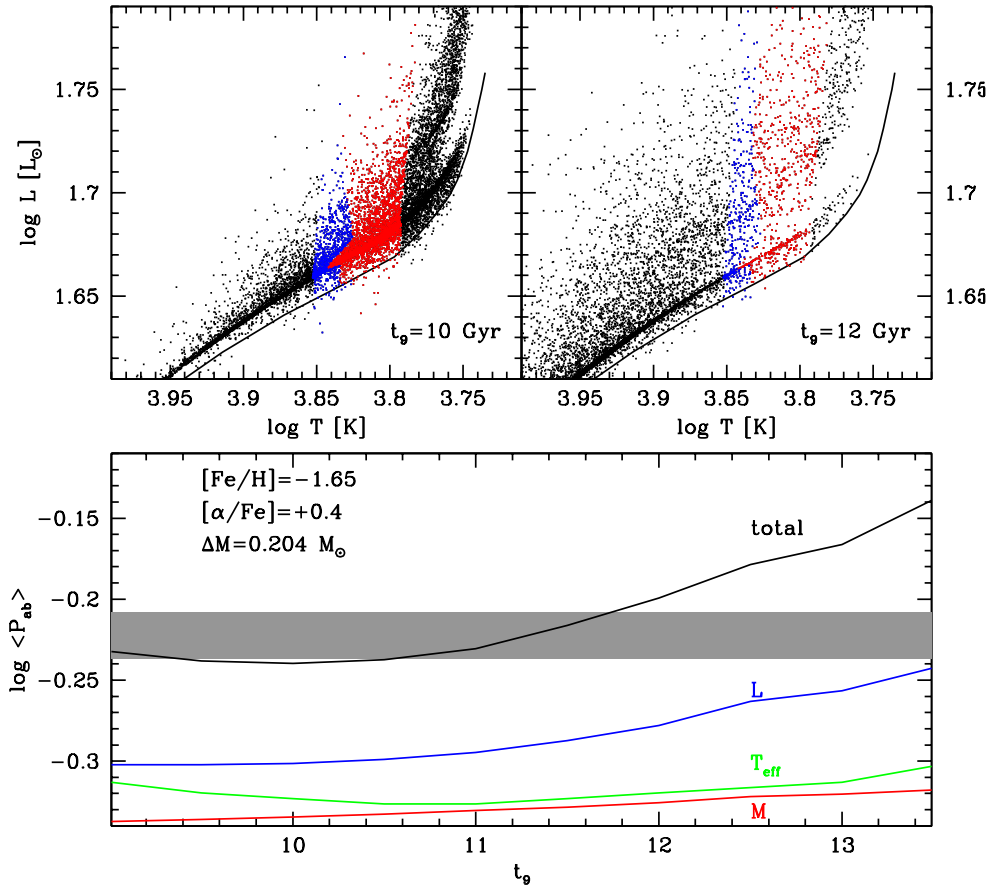}
 \caption{In the bottom panel, the contribution of each term of eq. \ref{per_eq}
 is shown as a function of the cluster age in a set of simulation made assuming
 [Fe/H]=-1.65, [$\alpha$/Fe]=+0.4 and a mass loss of $\Delta M=0.204~M_{\odot}$.
 The blue, red and green lines indicate the contribution of the luminosity, mass
 and temperature variation, respectively. The black line indicates the overall
 trend of the mean RR0 period given by the sum of the individual contributions.
 The grey shaded area indicates the Oosterhoff gap. In the top panels the
 synthetic HB of clusters simulated with the above mentioned metallicity and
 mass loss prescriptions and ages of 10 Gyr (top-left panel) and 12
 Gyr (top-right panel) are shown. Blue and red (pale and dark grey in the
 printed version of the paper) points mark the RR1 and RR0
 variables, respectively. The ZAHB is also shown as a black solid line.}
\label{ageeff}
\end{figure*}

\begin{figure*}
 \includegraphics[width=13cm]{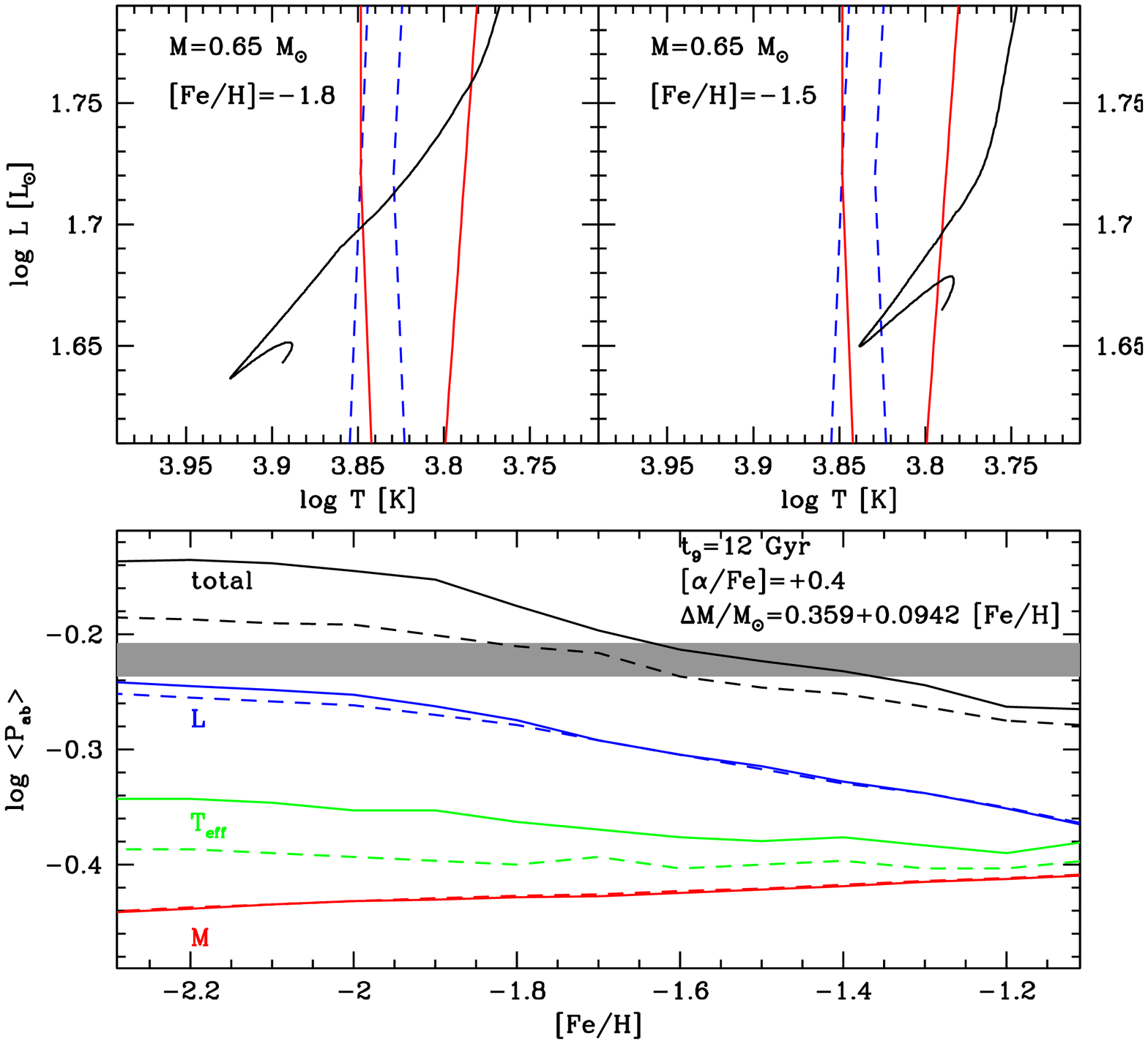}
 \caption{In the bottom panel, the contribution of each term of eq. \ref{per_eq}
 is shown as a function of the cluster metallicity in a set of simulation made 
 assuming
 $t_{9}$=12 Gyr, [$\alpha$/Fe]=+0.4 and a the mass loss-metallicity relation by
 Gratton et al. (2010). 
 The solid blue, red and green lines indicate the contributions of the luminosity, mass
 and temperature variation, respectively. The black line indicates the overall
 trend of the mean RR0 period given by the sum of the individual contributions.
 The dashed lines refer to the simulations where the hysteresis mechanism has
 been switched-off.
 The grey shaded area indicates the Oosterhoff gap. In the top panels the
 evolutionary tracks of a 0.65 $M_{\odot}$ HB star are shown for two different
 metallicities. The boundaries of the RR0 (solid lines) and RR1 (dashed 
 lines) instability strips are also shown.}
\label{hysteff}
\end{figure*}

The procedure described in Sect. \ref{met_sec} allows to reproduce the
morphology of the CMD of the target clusters by tuning their ages and mass loss
producing as output the synthetic period of the RR0 variables. 
In Fig. \ref{corr} the best-fit ages, amount of mass loss
(defined as the difference between the evolving mass at the tip of the Red Giant
Branch and the mean HB mass), HB type, mean HB masses, fraction of RR1 and
Galactocentric distances are plotted as a
function of the cluster metallicity. It is apparent that OoI and OoII clusters
occupy different regions of these diagrams. In particular:
\begin{itemize}
\item{All OoI clusters are younger than OoII ones;}
\item{All OoI clusters have redder HB morphologies;}
\item{All OoI clusters have a smaller fraction of RR1 than OoII ones;}
\item{OoI and OoII clusters follow a common mass loss-metallicity relation;}
\item{OoI clusters lie tendentially at larger distances from the Galactic
center with respect to OoII ones;}
\item{OoInt clusters are spread across all diagrams and do not follow any
clear trend with metallicity.}
\end{itemize} 
From the above considerations, a natural question arises: {\it can the above
correlations produce the Oosterhoff dichotomy?}.

In the top panel of Fig. \ref{res} the synthetic and observed mean RR0 periods are compared as
a function of metallicity. It is apparent that {\it the Oosterhoff dichotomy
among Galactic GCs is naturally reproduced by models}. In general, the
synthetic mean RR0 periods agree with the observed ones with differences smaller than
$\Delta \langle P_{ab}\rangle < 0.02~d$ although the mean RR0 periods of OoII
clusters appear to be systematically overestimated by $\sim 0.02~d$.
For reference, the mean difference between the average RR0 periods of OoI and
OoII clusters in this set of simulations (hereafter referred as the {\it bestfit} set) turns out to be $\Delta \langle P_{ab}\rangle=0.118~d$ which is
very close to the observed value ($\Delta \langle P_{ab}\rangle=0.101~d$).
In the top panel of Fig. \ref{res} the synthetic and observed relative 
fractions of RR1 (defined as $F_{c}=N_{RR1}/(N_{RR0}+N_{RR1})$) are also compared. Also in this
case, the models reproduce the trend defined by observations with the OoII
clusters holding a larger fraction of RR1, although the ratio between the RR1
fraction is slightly underestimated ($F_{c,OoII}/F_{c,OoI}=1.700$ instead of the 
observed $F_{c,OoII}/F_{c,OoI}=2.525$).

There are many physical processes that are effective in producing the above
difference in the mean RR0 periods. Here, we
concentrate on the three main mechanisms that have been proposed in previous
studies: age differences, mass loss-metallicity relation and hysteresis.
To evaluate the relative impact of each of the above effects, we repeated the
simulations assuming alternatively {\it i)} the same ages ($t_{9}$=12 Gyr) 
for all clusters, {\it ii)} the same amount of mass loss 
($\Delta M=0.2~M_{\odot}$), and {\it iii)} a switch-off of the hysteresis
criterion i.e. assuming that all RR Lyrae falling into the OR zone are RR0
regerdless of their previous pulsation mode.
The mean difference between the average RR0 periods of OoI and
OoII clusters in the above defined sets of simulations have been then calculated
and compared with that of the {\it bestfit} set. We note that a significant part
of the mean period differences disappears when the same age is assumed for all
clusters (case {\it i)}; with a resulting $\Delta \langle
P_{ab}\rangle=0.064~d$) and when the hysteresis mechanism is switched-off (case
{\it iii)}; $\Delta \langle P_{ab}\rangle=0.073~d$). It is worth noting that also the ratio of the
RR1 fractions in OoI and OoII clusters is significantly affected by these
changes: while $F_{c,OoII}/F_{c,OoI}=1.700$ in the {\it bestfit} set, it decreases to
1.255 in case {\it i)} and to 0.748 in case {\it iii)}. A negligible effect is
instead produced by the mass loss-metallicity relation (case {\it ii)}; $\Delta
\langle P_{ab}\rangle=0.120~d$; $F_{c,OoII}/F_{c,OoI}=2.142$).  

To understand the physical reason of the above
result, we decompose the effect on the 
mean RR0 periods in the contributions of
the mass, temperature and luminosity variation (according to eq. \ref{per_eq}) 
as a function of the cluster age
in a set of simulations conducted with a fixed metallicity ([Fe/H]=-1.65;
intermediate between OoI and OoII groups) and amount of mass loss ($\Delta
M=0.204~M_{\odot}$) (see Fig. \ref{ageeff}). It is apparent that a cluster with 
such a metallicity and
mass loss can belong to either the OoI or the OoII group depending on its age.
At ages $t_{9}>10.5$ Gyr, all the terms of eq. \ref{per_eq} lead to an increase of periods with age.
The strongest contribution to the mean period variation is that of the increase
of RR0 luminosity in old clusters. In fact, in older clusters stars reach the HB 
phase with a smaller mass populating the blue portion of the HB at high
temperatures. At relatively young ages HB stars cover a temperature range
which include the entire instability strip. In this case, most of the stars
become RR0 when their luminosity is close to that of the ZAHB in a region of the
CMD where they spend
the largest fraction of their HB evolution. When the cluster age exceeds a critical 
value ($t_{9}>10.5~Gyr$ at
[Fe/H]=-1.65) the largest fraction of HB stars start their evolution in the 
blue part of the HB and cross the instability
strip during their off-ZAHB evolution at high luminosity (see the top panels of
Fig. \ref{ageeff}). A less important contribution is due to the smaller
masses of RR0 in older clusters. 

It is interesting to note the non-monotonic
contribution of the temperature variation of RR0 variables to the mean period:
at ages younger than $\sim$11 Gyr an increase in the cluster age has the effect
of increasing the mean temperature of RR0 (with a consequent decreasing effect
on the period) as a direct consequence of the smaller HB mass.
However, at older ages the mean temperature of RR0 decreases contributing to
the increase of their mean periods. This is a consequence of the hysteresis
process: when the temperature of the star on the ZAHB is smaller than the FOBE
(in young/metal-rich clusters) it can become RR0 during its entire evolution at 
temperature hotter than the FRE, even when it crosses the OR region. 
On the other hand, if the HB evolution starts at temperature
hotter than the FBE (in old/metal-poor clusters) the star crosses the OR region 
from its blue border pulsating as a RR1. In this last case, it becomes an RR0 
only after the crossing of the FORE at low temperature. Therefore, the real 
transition between the RR0 and RR1 pulsation mode occurs at temperatures
anticorrelated with the ZAHB temperatures. For this reason the relative fraction
of RR1 is larger in OoII clusters. This effect is shown in Fig.
\ref{hysteff} where the variation of the mean RR0 period as a function of
metallicity is shown for the cases with and without hysteresis. It is evident
that the largest impact of hysteresis on the mean RR0 period is due to its effect on
the RR0 temperatures.  
 
\begin{figure}
 \includegraphics[width=8.7cm]{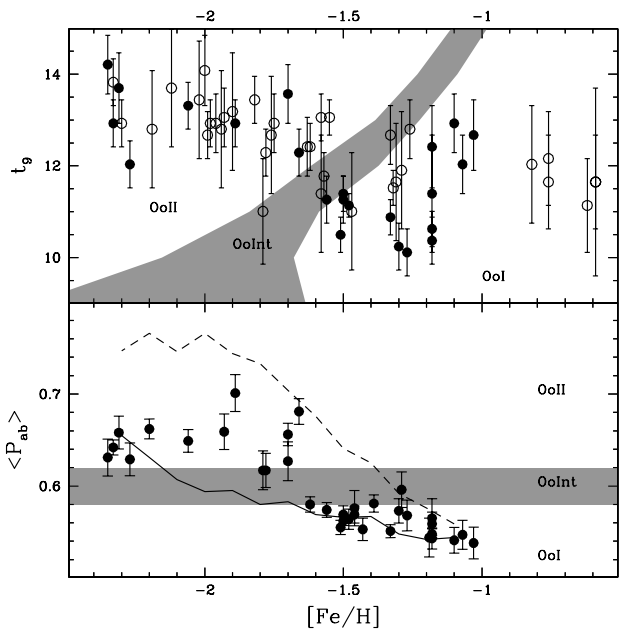}
 \caption{In the bottom panel, the $\langle P_{ab}\rangle$-[Fe/H] relations for
 clusters with $t_{9}$=10 Gyr (solid line) and $t_{9}$=13 Gyr (dashed line) 
 are shown for a set of simulations performed assuming the Gratton et al. (2010)
 mass loss-metallicity relation. The positions of GCs containing at least 10 RR0 is shown. 
 In the top panels the age-metallicity relation of Galactic GCs (from
 Mar{\'{\i}}n-Franch et al. 2009) is shown. Filled dots indicate GCs with at least
 10 RR0, open dots indicate GCs with less than 10 RR0. The region where OoInt clusters are
 expected to be located is shown by the grey shaded area.}
\label{agemet}
\end{figure}

Another interesting result of the analysis presented here is that all the OoI
and OoII clusters follow the same mass loss-metallicity relation, with the most
metal-poor clusters loosing a smaller fraction of mass. This evidence has been
already found by Gratton et al. (2010) who considered a large sample of GCs
spanning a wide range of metallicity. If we link the amount of 
mass loss to the cluster metallicity, and assume the same He, [$\alpha$/Fe] and 
CNO content, the morphology of the HB and the mean RR0 periods turn out 
to depend only on the cluster metallicity and age. In the bottom panel of 
Fig. \ref{agemet} the mean RR0 period-metallicity relations for two different
ages are shown. As expected, the younger is the cluster age the shorter is the
mean RR0 period. At all ages there is always a metallicity range where the
cluster cross the Oosterhoff gap. It is therefore possible to define a region in the
age-metallicity plane where OoI and OoII clusters reside. This is shown in the
top panel of Fig. \ref{agemet} where the GCs of the sample of Mar{\'{\i}}n-Franch et
al. (2009) are plotted. As already discussed by Mar{\'{\i}}n-Franch et al. (2009), GCs
distribute in the age-metallicity plane on two well-defined branches: one of them
is constituted by old ($t_{9}\sim 13$ Gyr) GCs spanning a wide range in
metallicity (the {\it old branch}), and the other constituted by GCs at larger distances from the
Galactic center following a steeper age-metallicity relation (the {\it young branch}). 
It can be noted that the region where OoInt
clusters are expected to reside constitutes a narrow strip intersecting the two
branches in a region almost devoid of GCs. In particular, the OoInt strip
crosses a gap in the metallicity distribution of the {\it old branch} at 
$-1.5<[Fe/H]<-1.3$\footnote{The same gap is present adopting the ages by
VandenBerg et al. (2013).}. Because of this gap, all the OoI clusters in this
metallicity range belong to the {\it young branch} and are generally at larger 
distances from the Galactic center. 
On the other hand, the slope of the age-metallicity 
relation of the {\it young branch} imply a very small intersection with the 
OoInt region (at $-1.6<[Fe/H]<-1.55$; $11<t_{9}<12$) where no GCs are present.

\subsection{Oosterhoff III: the case of NGC6388}
\label{ooiii_sec}

In this work we analysed the OoIII cluster NGC6388. In the assumption of a
canonical He abundance (Y=0.26), our simulation fail to reproduce
the mean RR0 period (with a predicted $\langle P_{ab}\rangle=0.468$,
significantly shorter than the observed value of $\langle P_{ab}\rangle=0.676$).
Moreover, to fit the anomalous morphology of the HB of this cluster we need to
assume a large amount of mass loss ($\Delta M=0.305~M_{\odot}$) which stray from
the mass loss-metallicity relation defined by the other clusters.
Many authors proposed a significantly larger He abundance for OoIII clusters to
account for the extended morphology of their HBs (Sweigart \& Catelan 1998; Busso et al. 2007; 
Yoon et al. 2008). An enhanced He abundance has the effect of increasing the
efficiency of the CNO cycle in the hydrogen-burning shell thus increasing the HB luminosity. Another effect is
to accelerate the stellar evolution so that He enhanced stars reach the HB phase
with a smaller mass with respect to He-normal stars of the same age. However, the
least massive stars are expected to populate the blue portion of the HB at
temperatures warmer than the instability strip. As a consequence, the net effect
of a He enhancement is the increase of periods (Marconi et al. 2011). We then
performed a set of simulation at different He abundances by interpolating across
the grid of He-enhanced tracks by BASTI database. The best-fit to the observed
mean RR0 period is obtained by adopting a He abundance of Y=0.35. This value is
in agreement with the amount needed to explain the extension and the slope
of the HB of this cluster (Busso et al. 2007). Another
effect of this assumption is the derived age: in case of a significant He 
enhancement, this
cluster turns out to be significantly younger than the other Milky Way GCs
($t_{9}\sim$7 Gyr). This is due to the fact that in He-rich stellar populations
the magnitude difference between the turn-off and the HB increases because of the
increased luminosity of the HB. Such a combination of He enhancement and young
age is however inconsistent with the observed magnitude difference of the RGB 
bump and the HB level (Raimondo et al. 2002). A possible solution to this
problem has been provided by Caloi \& D'Antona (2007) in an analogous analysis
in the other OoIII cluster NGC6441: they assumed the superposition of three stellar 
populations with
different He abundances, as already proposed for many GCs (Piotto 2009;
Carretta et al. 2009b) and observed in NGC6388 (Moretti et al. 2009; Bellini et al. 2013). 
Indeed, if 
the dominant cluster population of this
cluster would be He-poor it should have a red HB population which is not
expected to
produce a sizeable population of RR Lyrae, while the positions of its RGB bump and
turn-off would be compatible with that of an old GC. On the other hand a
fraction of He enhanced stars populate the region bracketing the instability
strip, thus producing the observed long mean RR0 period. 

\section{Extra-Galactic GCs: Oosterhoff Int}

We included in our analysis a sample of 4 old extra-Galactic GCs: one belonging to 
the LMC (NGC1466) and 3 belonging to the Fornax dSph (F2, F3 and F4). These
clusters have a metallicity comparable with that of the metal-poor OoII clusters
of the Milky Way but show mean RR0 periods comprised within the Oosterhoff gap.

The results of our simulations, conducted assuming a low $[\alpha/Fe]=0.0$, are 
shown in Fig. \ref{res}. It is noticeable
that our simulations overestimate the mean RR0 periods for 3 out 4 clusters
(with the only exception of F4). In particular, a systematic offset of $\Delta
\langle P_{ab}\rangle=0.046~d$ is noticeable for these 3 clusters. 
Also the relative fraction of RR1 are significantly overestimated.
Such 
differences, could be produced by many uncertain assumptions we made.

First, while the [$\alpha$/Fe] in these galaxies has been proved to be smaller
than what observed in the Milky Way at the same metallicity (Venn et al. 2004),
direct measurements of the [$\alpha$/Fe] abundance in two Fornax GCs of our sample (F2 and F3;
Letarte et al. 2006) showed an
$\alpha$-enhancement comparable with that observed among Galactic GCs. By repeating the simulations with the same
$\alpha$-enhanced mixture adopted for the Milky Way GCs, the difference between
the predicted and observed mean RR0 periods reduces to $\Delta
\langle P_{ab}\rangle=0.013~d$, which is comparable with the shift observed for
the Milky Way OoII GCs. 

Second, at least for Fornax GCs, Mackey \& Gilmore
(2003) reported that because of their short observational baseline both their
mean RR0 periods and the relative fraction of RR1 could be underestimated. 
Moreover, it is worth noting that spectroscopic
metallicities are not available for some of these clusters and in some case the
adopted [Fe/H] comes from photometric indices. However, in spite of these caution
notes, the observed discrepancy in the mean RR0 periods is associated to
an anomalously small amount of mass loss which stray from the relation defined by the
Milky Way GCs. A small amount of mass loss has been also reported by Salaris et al.
(2013) to reproduce the HB morphology of the Sculptor dSph. 
Note that in the present analysis the amount of mass loss is the
only free parameter which determines the HB morphology, so the difference in the
amount of mass loss simply reflects the differences in the HB morphology in
these GCs. 

In practice, as already noted by other authors (see e.g. Buonanno et al. 1998),
extra-Galactic GCs tends to show HB morphologies which are redder than those of
Galactic GCs with the same metallicity. 
Although the details of the mass loss process occurring during the
cluster lifetime are still uncertain, it is difficult to find a physical reason
why this stellar evolution process should depend on the host galaxy environment. 
It is therefore likely that the variation of other hidden parameters could produce 
the observed differences.

In analogy with what derived for OoIII cluster, a different He abundance can
significantly influence both the HB morphology and the mean RR0 periods. In this
case, a depletion of He of $\Delta Y\sim0.02$ is needed to reproduce the
observed mean RR0 periods and HB morphology implying an amount of mass loss
comparable to that of Milky Way GCs. However, our simulations have been already
performed assuming the cosmological He abundance (Y=0.245; Cyburt, Fields \&
Olive 2003; Salaris et al. 2004)
so that an unlikely sub-cosmological He abundance would be required for the 
extra-Galactic GCs to explain the observed differences, unless some deceptive
systematic uncertainties in the adopted theoretical framework are present. 

Another parameter affecting the HB
evolution is the abundance of the CNO elements. It has been indeed shown that an
enhanced CNO abundance produces an increase in luminosity and a decrease in 
the effective temperature of HB tracks (Dorman 1992). We then
performed a set of simulation assuming an $\alpha$-enhanced, CNO-enhanced
mixture ([$\alpha$/Fe]=+0.4; [CNO/Fe]=+0.3) by interpolating across
the grid of tracks by Pietrinferni et al. (2009). By adopting this
chemical composition, the
mean RR0 periods further decrease leading to a residual difference $\Delta
\langle P_{ab}\rangle=0.004~d$ with respect to observations. In this case the
amount of mass loss increase at values $\Delta M\sim0.18~M_{\odot}$, similar to 
those adopted for Milky Way GCs of the same metallicity.

\section{Conclusions}

In this paper we used CMD synthesis joined with the results of non-linear
pulsation models to simultaneously reproduce both the CMD morphology and the 
pulsational properties of RR0 variables for a sample of Galactic GCs which
determine the Oosterhoff dichotomy and for a sample of extra-Galactic GCs.
We find that the Oosterhoff dichotomy among Milky Way GCs is naturally reproduced by models with 
metal-poor OoII GCs ($[Fe/H]<-1.65$) having mean RR0 periods longer by $\Delta
P_{ab}\sim0.12~d$ with respect to OoI ones. A careful analysis of the relative
impact of the main physical processes affecting the mean RR0 periods in the
analysed GCs indicates that the observed difference is due to the combination of
two effects: {\it i)} the systematic age difference between metal-poor and
metal-rich GCs in the 
metallicity range $-1.9<[Fe/H]<-1.4$ (as already proposed by Lee \& Carney 1999), 
and {\it ii)} the hysteresis mechanism (van Albada \& Baker 1973).
The former process relies on the effect of the cluster age on the HB morphology
which favors a large fraction of evolved RR0 variables in those (old and/or 
metal-poor) GCs with a blue HB morphology (see Lee et al. 1990). The latter
process leads to a shift toward cold temperature (i.e. large periods) of the
effective border of the RR0/RR1 transition in OoII clusters. 

The analysed GCs belonging to the main Oosterhoff groups obey to the same mass
loss-metallicity relation, as already found by Gratton et al. (2010) while OoInt
and OoIII clusters stray from such a relation. In this picture, and assuming a
constant He, [$\alpha$/Fe] and CNO abundance, the mean RR0
periods of GCs depend univocally on their ages and metallicities.

It is interesting to note that, as a consequence of the above conclusions, the Oosterhoff
dichotomy can be viewed as a natural consequence of the lack of old
($t_{9}\sim 12$ Gyr) GCs in the metallicity range ($-1.5<[Fe/H]<-1.3$),
and the concomitant absence of young ($t_{9}<11$ Gyr) GCs at $[Fe/H]<-1.6$. It is 
not clear
whether these gaps in the age-metallicity relation of Galactic GCs are due to
stochastic effects or to a physical process linked to the mechanism of formation
and chemical evolution of the Galactic halo. 
In this regard, we note that these gaps are located close to the separation
between the two branches of the GCs age-metallicity relation. Leaman, VandenBerg
\& Mendel (2013)
proposed that GCs belonging to the {\it young branch} of the age-metallicity
relation could form in satellite galaxies later accreted by the Milky Way during
the early building-up process of the Galactic halo while those belonging to the
{\it old branch} formed in-situ.

At odds with what happens in Galactic GCs, the analysed GCs belonging to the 
nearby dSph (LMC and Fornax) exhibit shorter mean RR0 periods and redder HB
morphology with respect to Milky Way GCs of the same metallicity. 
This effect is compatible with the correlation between the integrated 
ultraviolet colors of GCs and the mass of their host galaxy found by Dalessandro
et al. (2012). An
investigation on the effect of the various input parameters indicates as a
possible solution a depleted He abundance or an increased [CNO/Fe] ratio in
these clusters. The presence of such chemical differences between Milky Way and
dSph GCs could be linked to their different chemical evolution.
In principle, the $\Delta Y/\Delta Z$ gradient is not expected to significantly
change in galaxies of different sizes (Balser 2006). However, differences in the mean He
content could be linked to the different GCs formation process in these galaxies. 
Indeed, the proposed scenario for the origin of the multiple stellar populations
in GCs predicts that a first
generation of stars polluted the intra-cluster medium where a more concentrated
second generation formed. The first generation then evaporates 
because of the interaction of the cluster with the host galaxy (D'Ercole et al. 2008).
The present-day fraction of first/second generation should therefore reflect the
strength of the tidal field which is expected to be smaller in dSph than in the
Milky Way (D'Antona et al. 2013). 
It has been shown that the chemical signatures of multiple population are
present among the HB stars 
of many Milky Way GCs (Marino et al. 2011; Gratton et al. 2011, 2012a,
2013, 2014). It is not clear if the same chemical anomalies (e.g. the Na-O 
anticorrelation) are present in extra-Galactic GCs (Johnson, Ivans \& Stetson 
2006; Mucciarelli et al. 2009; Mucciarelli, Origlia \& Ferraro 2010; Mateluna et
al. 2012). If
extra-Galactic GCs would contain a smaller (if any) fraction of second
generation He-enhanced
stars with respect to Galactic GCs, this would explain a difference in their 
average He content. This however should involve a problem in the zero point of
the mean RR0 periods which appear to be consistent with a cosmological He
abundance in Galactic GCs. Although the mean RR0 period is very sensitive to the
uncertainties in the temperature of the FRE (mainly linked to the uncertainty on the convection
theory) and in the HB luminosity, this hypothesis requires a
deeper investigation. A CNO enhancement can produce a similar effect. In
principle dSph, because of their shallow potential well, are expected to retain
a small fraction of elements produced in SNe explosions (like Fe) with respect
to those produced during the Asymptotic Giant Branch nucleosynthesis (like C, N
and O). So, a larger [CNO/Fe] abundance pattern could be present in these
clusters. Unfortunately, reliable measures of the C+N+O abundance are now
available only for few Galactic GCs, often providing conflicting results 
(Yong et al. 2009; Villanova, Geisler \& Piotto 2010; Gratton et al. 2012b).

The comparison of the predicted and observed RR0 properties of NGC6388, the only
OoIII cluster analysed here, suggests that the RR0 population of this cluster
must be constituted by He-enhanced ($Y\sim0.35$) stars. This conclusion is in
agreement with what already proposed for OoIII clusters by many authors on the
basis of its HB morphology (Sweigart \& Catelan 1998; Moehler \&
Sweigart 2006; Yoon et al. 2008). However, considerations on the 
magnitude difference between the HB, the RGB bump and the turn-off of this cluster 
imply that the dominant cluster population must have a canonical He abundance
and does not contribute to the RR Lyrae population of this cluster. This is
consistent with the evidence of multiple stellar populations observed in this
GC (Moretti et al. 2009; Bellini et al. 2013) and in many other Galactic GCs.

\section*{Acknowledgments}

This research has been founded by PRIN INAF 2011 "Multiple populations in globular
clusters: their role in the Galaxy assembly" (PI E. Carretta) and PRIN MIUR 
2010-2011 "The Chemical and Dynamical Evolution of the Milky Way and Local Group Galaxies" (PI F. Matteucci).
GF has been supported by Futuro in Ricerca 2013 (RBFR13J716). We warmly thank
Marcella Marconi for the useful discussion and the anonymous referee for his/her
comments and suggestions.

\label{lastpage}


\begin{thebibliography}{99}

\bibitem[\protect\citeauthoryear{Alves-Brito et 
al.}{2012}]{2012A&A...540A...3A} Alves-Brito A., Yong D., Mel{\'e}ndez J., V{\'a}squez S., Karakas A.~I., 2012, A\&A, 540, A3 
\bibitem[\protect\citeauthoryear{Ashman, Bird, 
\& Zepf}{1994}]{1994AJ....108.2348A} Ashman K.~M., Bird C.~M., Zepf S.~E., 1994, AJ, 108, 2348 
\bibitem[\protect\citeauthoryear{Balser}{2006}]{2006AJ....132.2326B} Balser 
D.~S., 2006, AJ, 132, 2326 
\bibitem[\protect\citeauthoryear{Bellini et 
al.}{2013}]{2013ApJ...765...32B} Bellini A., et al., 2013, ApJ, 765, 32 
\bibitem[\protect\citeauthoryear{Bono, Caputo, 
\& Marconi}{1995}]{1995AJ....110.2365B} Bono G., Caputo F., Marconi M., 1995, AJ, 110, 2365 
\bibitem[\protect\citeauthoryear{Bono et al.}{1995}]{1995ApJ...448L.115B} 
Bono G., Caputo F., Castellani V., Marconi M., 1995, ApJ, 448, L115 
\bibitem[\protect\citeauthoryear{Buonanno et 
al.}{1990}]{1990AJ....100.1811B} Buonanno R., Buscema G., Fusi Pecci F., 
Richer H.~B., Fahlman G.~G., 1990, AJ, 100, 1811 
\bibitem[\protect\citeauthoryear{Buonanno et 
al.}{1999}]{1999AJ....118.1671B} Buonanno R., Corsi C.~E., Castellani M., 
Marconi G., Fusi Pecci F., Zinn R., 1999, AJ, 118, 1671 
\bibitem[\protect\citeauthoryear{Buonanno et 
al.}{1998}]{1998ApJ...501L..33B} Buonanno R., Corsi C.~E., Zinn R., Fusi 
Pecci F., Hardy E., Suntzeff N.~B., 1998, ApJ, 501, L33 
\bibitem[\protect\citeauthoryear{Busso et 
al.}{2007}]{2007A&A...474..105B} Busso G., et al., 2007, A\&A, 474, 105 
\bibitem[\protect\citeauthoryear{Caloi 
\& D'Antona}{2007}]{2007A&A...463..949C} Caloi V., D'Antona F., 2007, A\&A, 463, 949 
\bibitem[\protect\citeauthoryear{Caputo, Tornambe, 
\& Castellani}{1978}]{1978A&A....67..107C} Caputo F., Tornambe A., Castellani V., 1978, A\&A, 67, 107 
\bibitem[\protect\citeauthoryear{Carretta et 
al.}{2009}]{2009A&A...508..695C} Carretta E., Bragaglia A., Gratton R., D'Orazi
V., Lucatello S., 2009a, A\&A, 508, 695 
\bibitem[\protect\citeauthoryear{Carretta et 
al.}{2009}]{2009A&A...505..117C} Carretta E., et al., 2009b, A\&A, 505, 117 
\bibitem[\protect\citeauthoryear{Cassisi et 
al.}{2004}]{2004A&A...426..641C} Cassisi S., Castellani M., Caputo F., Castellani V., 2004, A\&A, 426, 641 
\bibitem[\protect\citeauthoryear{Cassisi et 
al.}{1998}]{1998A&AS..129..267C} Cassisi S., Castellani V., degl'Innocenti S., Weiss A., 1998, A\&AS, 129, 267 
\bibitem[\protect\citeauthoryear{Castellani, Caputo, 
\& Castellani}{2003}]{2003A&A...410..871C} Castellani M., Caputo F., Castellani V., 2003, A\&A, 410, 871 
\bibitem[\protect\citeauthoryear{Castellani, Castellani, 
\& Cassisi}{2005}]{2005A&A...437.1017C} Castellani M., Castellani V., Cassisi S., 2005, A\&A, 437, 1017 
\bibitem[\protect\citeauthoryear{Castellani 
\& degl'Innocenti}{1999}]{1999A&A...344...97C} Castellani V., degl'Innocenti S., 1999, A\&A, 344, 97 
\bibitem[\protect\citeauthoryear{Catelan}{2009}]{2009Ap&SS.320..261C} Catelan M., 2009, Ap\&SS, 320, 261 
\bibitem[\protect\citeauthoryear{Clement et 
al.}{2001}]{2001AJ....122.2587C} Clement C.~M., et al., 2001, AJ, 122, 2587 
\bibitem[\protect\citeauthoryear{Clement 
\& Shelton}{1999}]{1999ApJ...515L..85C} Clement C.~M., Shelton I., 1999, ApJ, 515, L85 
\bibitem[\protect\citeauthoryear{Coppola et 
al.}{2013}]{2013ApJ...775....6C} Coppola G., et al., 2013, ApJ, 775, 6 
\bibitem[\protect\citeauthoryear{Cseresnjes}{2001}]{2001A&A...375..909C} Cseresnjes P., 2001, A\&A, 375, 909 
\bibitem[\protect\citeauthoryear{Cusano et al.}{2013}]{2013ApJ...779....7C} 
Cusano F., et al., 2013, ApJ, 779, 7 
\bibitem[\protect\citeauthoryear{Cyburt, Fields, 
\& Olive}{2003}]{2003PhLB..567..227C} Cyburt R.~H., Fields B.~D., Olive K.~A., 2003, PhLB, 567, 227 
\bibitem[\protect\citeauthoryear{Dalessandro et 
al.}{2012}]{2012AJ....144..126D} Dalessandro E., Schiavon R.~P., Rood 
R.~T., Ferraro F.~R., Sohn S.~T., Lanzoni B., O'Connell R.~W., 2012, AJ, 
144, 126 
\bibitem[\protect\citeauthoryear{Dall'Ora et 
al.}{2003}]{2003AJ....126..197D} Dall'Ora M., et al., 2003, AJ, 126, 197 
\bibitem[\protect\citeauthoryear{D'Antona et 
al.}{2013}]{2013MNRAS.434.1138D} D'Antona F., Caloi V., D'Ercole A., Tailo 
M., Vesperini E., Ventura P., Di Criscienzo M., 2013, MNRAS, 434, 1138 
\bibitem[\protect\citeauthoryear{D'Ercole et 
al.}{2008}]{2008MNRAS.391..825D} D'Ercole A., Vesperini E., D'Antona F., 
McMillan S.~L.~W., Recchi S., 2008, MNRAS, 391, 825 
\bibitem[\protect\citeauthoryear{di Criscienzo et 
al.}{2011}]{2011MNRAS.414.3381D} di Criscienzo M., et al., 2011, MNRAS, 
414, 3381 
\bibitem[\protect\citeauthoryear{Dorman}{1992}]{1992ApJS...80..701D} Dorman 
B., 1992, ApJS, 80, 701 
\bibitem[\protect\citeauthoryear{{\'E}igenson 
\& Yatsyk}{1985}]{1985Ap.....22..246E} {\'E}igenson A.~M., Yatsyk O.~S., 1985, Ap, 22, 246 
\bibitem[\protect\citeauthoryear{Fiorentino et 
al.}{2012}]{2012ApJ...759L..12F} Fiorentino G., Stetson P.~B., Monelli M., 
Bono G., Bernard E.~J., Pietrinferni A., 2012, ApJ, 759, L12 
\bibitem[\protect\citeauthoryear{Garofalo et 
al.}{2013}]{2013ApJ...767...62G} Garofalo A., et al., 2013, ApJ, 767, 62 
\bibitem[\protect\citeauthoryear{Gratton et 
al.}{2010}]{2010A&A...517A..81G} Gratton R.~G., Carretta E., Bragaglia A., Lucatello S., D'Orazi V., 2010, A\&A, 517, A81 
\bibitem[\protect\citeauthoryear{Gratton et 
al.}{2011}]{2011A&A...534A.123G} Gratton R.~G., Lucatello S., Carretta E., Bragaglia A., D'Orazi V., Momany Y.~A., 2011, A\&A, 534, A123 
\bibitem[\protect\citeauthoryear{Gratton et 
al.}{2012}]{2012A&A...539A..19G} Gratton R.~G., et al., 2012a, A\&A, 539, A19 
\bibitem[\protect\citeauthoryear{Gratton et 
al.}{2012}]{2012A&A...544A..12G} Gratton R.~G., Villanova S., Lucatello S.,
Sollima A., Geisler D., Carretta E., Cassisi S., Bragaglia A., 2012b, A\&A, 544, A12 
\bibitem[\protect\citeauthoryear{Gratton et 
al.}{2013}]{2013A&A...549A..41G} Gratton R.~G., et al., 2013, A\&A, 549, A41 
\bibitem[\protect\citeauthoryear{Gratton et 
al.}{2014}]{2014A&A...563A..13G} Gratton R.~G., et al., 2014, A\&A, 563, A13 
\bibitem[\protect\citeauthoryear{Greco et al.}{2007}]{2007ApJ...670..332G} 
Greco C., et al., 2007, ApJ, 670, 332 
\bibitem[\protect\citeauthoryear{Greco et al.}{2009}]{2009ApJ...701.1323G} 
Greco C., et al., 2009, ApJ, 701, 1323 
\bibitem[\protect\citeauthoryear{Grosse}{1932}]{grosse} 
Grosse E., 1932, AN, 246, 402 
\bibitem[\protect\citeauthoryear{Hachenberg}{1939}]{hachenberg} 
Hachenberg O., 1939, ZA, 18, 49 
\bibitem[\protect\citeauthoryear{Jang et al.}{2014}]{2014arXiv1404.7508J} 
Jang S., Lee Y.-W., Joo S.-J., Na C., 2014, MNRAS, in press, arXiv:1404.7508 
\bibitem[\protect\citeauthoryear{Johnson et 
al.}{1999}]{1999ApJ...527..199J} Johnson J.~A., Bolte M., Stetson P.~B., 
Hesser J.~E., Somerville R.~S., 1999, ApJ, 527, 199 
\bibitem[\protect\citeauthoryear{Johnson, Ivans, 
\& Stetson}{2006}]{2006ApJ...640..801J} Johnson J.~A., Ivans I.~I., Stetson P.~B., 2006, ApJ, 640, 801 
\bibitem[\protect\citeauthoryear{Kinemuchi et 
al.}{2008}]{2008AJ....136.1921K} Kinemuchi K., Harris H.~C., Smith H.~A., 
Silbermann N.~A., Snyder L.~A., La Cluyz{\'e} A.~P., Clark C.~L., 2008, AJ, 
136, 1921 
\bibitem[\protect\citeauthoryear{Kinman}{1959}]{1959MNRAS.119..134K} Kinman 
T.~D., 1959, MNRAS, 119, 134 
\bibitem[\protect\citeauthoryear{Kroupa}{2002}]{2002Sci...295...82K} Kroupa 
P., 2002, Sci, 295, 82 
\bibitem[\protect\citeauthoryear{Kuehn et al.}{2008}]{2008ApJ...674L..81K} 
Kuehn C., et al., 2008, ApJ, 674, L81 
\bibitem[\protect\citeauthoryear{Leaman, VandenBerg, 
\& Mendel}{2013}]{2013MNRAS.436..122L} Leaman R., VandenBerg D.~A., Mendel J.~T., 2013, MNRAS, 436, 122 
\bibitem[\protect\citeauthoryear{Lee}{1991}]{1991ApJ...367..524L} Lee 
Y.-W., 1991, ApJ, 367, 524 
\bibitem[\protect\citeauthoryear{Lee 
\& Carney}{1999}]{1999AJ....118.1373L} Lee J.-W., Carney B.~W., 1999, AJ, 118, 1373 
\bibitem[\protect\citeauthoryear{Lee, Demarque, 
\& Zinn}{1990}]{1990ApJ...350..155L} Lee Y.-W., Demarque P., Zinn R., 1990, ApJ, 350, 155 
\bibitem[\protect\citeauthoryear{Lee, Demarque, 
\& Zinn}{1994}]{1994ApJ...423..248L} Lee Y.-W., Demarque P., Zinn R., 1994, ApJ, 423, 248 
\bibitem[\protect\citeauthoryear{Lemasle et 
al.}{2013}]{2013A&A...558A..31L} Lemasle B., et al., 2013, A\&A, 558, A31 
\bibitem[\protect\citeauthoryear{Letarte et 
al.}{2006}]{2006A&A...453..547L} Letarte B., Hill V., Jablonka P., Tolstoy E., Fran{\c c}ois P., Meylan G., 2006, A\&A, 453, 547 
\bibitem[\protect\citeauthoryear{Marconi et 
al.}{2011}]{2011ApJ...738..111M} Marconi M., Bono G., Caputo F., Piersimoni 
A.~M., Pietrinferni A., Stellingwerf R.~F., 2011, ApJ, 738, 111 
\bibitem[\protect\citeauthoryear{Marconi et 
al.}{2003}]{2003ApJ...596..299M} Marconi M., Caputo F., Di Criscienzo M., 
Castellani M., 2003, ApJ, 596, 299 
\bibitem[\protect\citeauthoryear{Mar{\'{\i}}n-Franch et 
al.}{2009}]{2009ApJ...694.1498M} Mar{\'{\i}}n-Franch A., et al., 2009, ApJ, 
694, 1498 
\bibitem[\protect\citeauthoryear{Marino et al.}{2011}]{2011ApJ...730L..16M} 
Marino A.~F., Villanova S., Milone A.~P., Piotto G., Lind K., Geisler D., 
Stetson P.~B., 2011, ApJ, 730, L16 
\bibitem[\protect\citeauthoryear{Marino et 
al.}{2012}]{2012A&A...541A..15M} Marino A.~F., et al., 2012, A\&A, 541, A15 
\bibitem[\protect\citeauthoryear{Mateluna et 
al.}{2012}]{2012A&A...548A..82M} Mateluna R., Geisler D., Villanova S., Carraro G., Grocholski A., Sarajedini A., Cole A., Smith V., 2012, A\&A, 548, A82 
\bibitem[\protect\citeauthoryear{Miceli et al.}{2008}]{2008ApJ...678..865M} 
Miceli A., et al., 2008, ApJ, 678, 865 
\bibitem[\protect\citeauthoryear{Milone et al.}{2014}]{2014ApJ...785...21M} 
Milone A.~P., et al., 2014, ApJ, 785, 21 
\bibitem[\protect\citeauthoryear{Moehler 
\& Sweigart}{2006}]{2006A&A...455..943M} Moehler S., Sweigart A.~V., 2006, A\&A, 455, 943 
\bibitem[\protect\citeauthoryear{Moretti et 
al.}{2009}]{2009A&A...493..539M} Moretti A., et al., 2009, A\&A, 493, 539 
\bibitem[\protect\citeauthoryear{Mucciarelli, Origlia, 
\& Ferraro}{2010}]{2010ApJ...717..277M} Mucciarelli A., Origlia L., Ferraro F.~R., 2010, ApJ, 717, 277 
\bibitem[\protect\citeauthoryear{Mucciarelli et 
al.}{2009}]{2009ApJ...695L.134M} Mucciarelli A., Origlia L., Ferraro F.~R., 
Pancino E., 2009, ApJ, 695, L134 
\bibitem[\protect\citeauthoryear{Norris}{2004}]{2004ApJ...612L..25N} Norris 
J.~E., 2004, ApJ, 612, L25 
\bibitem[\protect\citeauthoryear{Oosterhoff}{1939}]{1939Obs....62..104O} 
Oosterhoff P.~T., 1939, Obs, 62, 104 
\bibitem[\protect\citeauthoryear{Oosterhoff}{1944}]{1944BAN....10...55O} 
Oosterhoff P.~T., 1944, BAN, 10, 55 
\bibitem[\protect\citeauthoryear{Pietrinferni et 
al.}{2006}]{2006ApJ...642..797P} Pietrinferni A., Cassisi S., Salaris M., 
Castelli F., 2006, ApJ, 642, 797 
\bibitem[\protect\citeauthoryear{Pietrinferni et 
al.}{2009}]{2009ApJ...697..275P} Pietrinferni A., Cassisi S., Salaris M., 
Percival S., Ferguson J.~W., 2009, ApJ, 697, 275 
\bibitem[\protect\citeauthoryear{Piotto}{2009}]{2009IAUS..258..233P} Piotto 
G., 2009, IAUS, 258, 233 
\bibitem[\protect\citeauthoryear{Piotto et 
al.}{2002}]{2002A&A...391..945P} Piotto G., et al., 2002, A\&A, 391, 945 
\bibitem[\protect\citeauthoryear{Piotto et al.}{2007}]{2007ApJ...661L..53P} 
Piotto G., et al., 2007, ApJ, 661, L53 
\bibitem[\protect\citeauthoryear{Pritzl et al.}{2000}]{2000ApJ...530L..41P} 
Pritzl B., Smith H.~A., Catelan M., Sweigart A.~V., 2000, ApJ, 530, L41 
\bibitem[\protect\citeauthoryear{Raimondo et 
al.}{2002}]{2002ApJ...569..975R} Raimondo G., Castellani V., Cassisi S., 
Brocato E., Piotto G., 2002, ApJ, 569, 975 
\bibitem[\protect\citeauthoryear{Rich et al.}{1997}]{1997ApJ...484L..25R} 
Rich R.~M., et al., 1997, ApJ, 484, L25 
\bibitem[\protect\citeauthoryear{Rosenberg et 
al.}{2000}]{2000A&AS..145..451R} Rosenberg A., Aparicio A., Saviane I., Piotto
G., 2000a, A\&AS, 145, 451 
\bibitem[\protect\citeauthoryear{Rosenberg et 
al.}{2000}]{2000A&AS..144....5R} Rosenberg A., Piotto G., Saviane I., Aparicio
A., 2000b, A\&AS, 144, 5 
\bibitem[\protect\citeauthoryear{Salaris, Cassisi, 
\& Pietrinferni}{2008}]{2008ApJ...678L..25S} Salaris M., Cassisi S., Pietrinferni A., 2008, ApJ, 678, L25 
\bibitem[\protect\citeauthoryear{Salaris et 
al.}{2013}]{2013A&A...559A..57S} Salaris M., de Boer T., Tolstoy E., Fiorentino G., Cassisi S., 2013, A\&A, 559, A57 
\bibitem[\protect\citeauthoryear{Salaris et 
al.}{2004}]{2004A&A...420..911S} Salaris M., Riello M., Cassisi S., Piotto G., 2004, A\&A, 420, 911 
\bibitem[\protect\citeauthoryear{Sandage}{1957}]{sandage} 
Sandage A., 1957, in "Stellar populations", ed. D. O'Connell, Amsterdam, 
p. 41
\bibitem[\protect\citeauthoryear{Sandage}{1969}]{1969ApJ...157..515S} 
Sandage A., 1969, ApJ, 157, 515 
\bibitem[\protect\citeauthoryear{Sandage}{1981}]{1981ApJ...248..161S} 
Sandage A., 1981, ApJ, 248, 161 
\bibitem[\protect\citeauthoryear{Sandage}{1982}]{1982ApJ...252..553S} 
Sandage A., 1982b, ApJ, 252, 553 
\bibitem[\protect\citeauthoryear{Sandage}{1982}]{1982ApJ...252..574S} 
Sandage A., 1982a, ApJ, 252, 574 
\bibitem[\protect\citeauthoryear{Sandage}{1990}]{1990ApJ...350..631S} 
Sandage A., 1990, ApJ, 350, 631 
\bibitem[\protect\citeauthoryear{Sandage, Katem, 
\& Sandage}{1981}]{1981ApJS...46...41S} Sandage A., Katem B., Sandage M., 1981, ApJS, 46, 41 
\bibitem[\protect\citeauthoryear{Sawyer Hogg}{1973}]{1973PDDO....3....6S} 
Sawyer Hogg H., 1973, PDDO, 3, 6 
\bibitem[\protect\citeauthoryear{Schwarzchild}{1957}]{swarzschild} 
Schwarzschild M., 1957, in "Stellar populations", ed. D. O'Connell, Amsterdam, 
p. 59 
\bibitem[\protect\citeauthoryear{Siegel 
\& Majewski}{2000}]{2000AJ....120..284S} Siegel M.~H., Majewski S.~R., 2000, AJ, 120, 284 
\bibitem[\protect\citeauthoryear{Soszynski et 
al.}{2003}]{2003AcA....53...93S} Soszynski I., et al., 2003, AcA, 53, 93 
\bibitem[\protect\citeauthoryear{Stellingwerf}{1975}]{1975ApJ...195..441S} 
Stellingwerf R.~F., 1975, ApJ, 195, 441 
\bibitem[\protect\citeauthoryear{Stobie}{1971}]{1971ApJ...168..381S} Stobie 
R.~S., 1971, ApJ, 168, 381 
\bibitem[\protect\citeauthoryear{Suntzeff, Kinman, 
\& Kraft}{1991}]{1991ApJ...367..528S} Suntzeff N.~B., Kinman T.~D., Kraft R.~P., 1991, ApJ, 367, 528 
\bibitem[\protect\citeauthoryear{Sweigart 
\& Catelan}{1998}]{1998ApJ...501L..63S} Sweigart A.~V., Catelan M., 1998, ApJ, 501, L63 
\bibitem[\protect\citeauthoryear{Szczygie{\l}, Pojma{\'n}ski, 
\& Pilecki}{2009}]{2009AcA....59..137S} Szczygie{\l} D.~M., Pojma{\'n}ski G., Pilecki B., 2009, AcA, 59, 137 
\bibitem[\protect\citeauthoryear{van Albada 
\& Baker}{1973}]{1973ApJ...185..477V} van Albada T.~S., Baker N., 1973, ApJ, 185, 477 
\bibitem[\protect\citeauthoryear{VandenBerg et 
al.}{2013}]{2013ApJ...775..134V} VandenBerg D.~A., Brogaard K., Leaman R., 
Casagrande L., 2013, ApJ, 775, 134 
\bibitem[\protect\citeauthoryear{Venn et al.}{2004}]{2004AJ....128.1177V} 
Venn K.~A., Irwin M., Shetrone M.~D., Tout C.~A., Hill V., Tolstoy E., 
2004, AJ, 128, 1177 
\bibitem[\protect\citeauthoryear{Villanova et 
al.}{2013}]{2013ApJ...778..186V} Villanova S., Geisler D., Carraro G., Moni 
Bidin C., Mu{\~n}oz C., 2013, ApJ, 778, 186 
\bibitem[\protect\citeauthoryear{Villanova, Geisler, 
\& Piotto}{2010}]{2010ApJ...722L..18V} Villanova S., Geisler D., Piotto G., 2010, ApJ, 722, L18 
\bibitem[\protect\citeauthoryear{Walker}{1992}]{1992AJ....104.1395W} Walker 
A.~R., 1992, AJ, 104, 1395 
\bibitem[\protect\citeauthoryear{Yoon et al.}{2008}]{2008ApJ...677.1080Y} 
Yoon S.-J., Joo S.-J., Ree C.~H., Han S.-I., Kim D.-G., Lee Y.-W., 2008, 
ApJ, 677, 1080 
\bibitem[\protect\citeauthoryear{Yong et al.}{2009}]{2009ApJ...695L..62Y} 
Yong D., Grundahl F., D'Antona F., Karakas A.~I., Lattanzio J.~C., Norris 
J.~E., 2009, ApJ, 695, L62 
\bibitem[\protect\citeauthoryear{Zinn}{1986}]{1986ApJ...304..579Z} Zinn R., 
1986, ApJ, 304, 579 
\bibitem[\protect\citeauthoryear{Zorotovic et 
al.}{2009}]{2009AJ....137..257Z} Zorotovic M., Catelan M., Zoccali M., 
Pritzl B.~J., Smith H.~A., Stephens A.~W., Contreras R., Escobar M.~E., 
2009, AJ, 137, 257 

\end{thebibliography}
\end{document}